# Optical properties differences across the Insulator-Metal Transition in VO$_2$ Thin Films grown on different substrates

E. Radue,[1,*] E. Crisman,[1] L. Wang,[1] S. Kittiwatanakul,[2] J. Lu,[3] S. A. Wolf,[2, 3] R. Wincheski,[4] R. A. Lukaszew,[1] and I. Novikova[1]

[1]*Department of Physics, College of William and Mary, Williamsburg, VA 23187, USA*
[2]*Department of Physics, University of Virginia, Charlottesville, VA 22904, USA*
[3]*Department of Material Sciences and Engineering, University of Virginia, Charlottesville, VA 22904, USA*
[4]*NASA Langley Research Center, Hampton, VA 23681 USA*
*Corresponding author: elradue@email.wm.edu

Abstract:

We have used Raman spectroscopy to investigate the optical properties of vanadium dioxide (VO$_2$) thin films deposited on different substrates during the thermally induced insulating to metallic phase transition. We observed a significant difference in transition temperature in VO$_2$ films similarly grown on quartz and sapphire substrates: the film grown on quartz displayed the phase transition at a lower temperature ($T_c$=50 $^o$C) compared a film grown on sapphire ($T_c$=68 $^o$C). We also investigated differences in the detected Raman signal for different wavelengths and polarizations of the excitation laser. We found that for either substrate, a longer wavelength (in our case 785 nm) yielded the clearest VO$_2$ Raman spectra, with no polarization dependence.

I.  Introduction

Vanadium dioxide (VO$_2$) undergoes a well-known thermally induced phase transition, changing from an insulator with a monoclinic lattice structure to a conductor with a tetragonal lattice structure [1- 3]. This structural insulator-to-metal transition (IMT) is evidenced by strong changes in the electrical and optical properties between the two VO$_2$ phases. The fact that MIT in VO$_2$ occurs just above room temperature makes it an exciting prospect for many new technologies. For example, it offers a lower loss alternative to noble metals for plasmonic applications [4], development of smart window coatings [5, 6], novel electronic devices [7,8], etc. Also, the MIT can be optically induced in a sub-ps timescale [9] making it an attractive material for ultrafast optical switches and sensors.

Many proposed applications would require deposition of vanadium dioxide thin films on a variety of suitable substrates and a comparison of substrate effects is needed for specific cases. Thus, in this paper we experimentally investigate the effects of different substrates by comparing the optical properties of VO$_2$ films grown on c-Al$_2$O$_3$ (sapphire) and crystalline SiO$_2$ (quartz) substrates. We have monitored the changes in the VO$_2$ structure using Raman spectroscopy. The two phases of VO$_2$ have distinct Raman signatures: the insulating monoclinic phase (at low temperature) displays several sharp Raman peaks, and at temperatures above the MIT the Raman spectrum of the metallic rutile phase is characterized by a broadband emission [13]. While we observed this typical behavior in all samples, this transition occurred at a lower temperature for the samples grown on quartz, compared to those grown on sapphire. Complementary measurements of the optical reflectivity showed a similar trend. This change in the apparent transition temperature is likely to be caused by different film microstructure, since different crystalline substrates can lead to epitaxial growth while others may not. In turn epitaxial growth can involve strain due to lattice mismatches while polycrystalline films may alter nucleation dynamics at grain boundaries. We also experimentally found an optimal wavelength for the excitation laser leading to optimal VO$_2$ Raman spectra, and found that the near-IR (785 nm) laser yielded the best resolution.

II. Sample characterization and experimental Set Up

The VO$_2$ thin films in this study were prepared using Reactive Biased Target Ion Beam Deposition (RBTIBD), which is described at length elsewhere [12]. The microstructure of the VO$_2$ thin films was characterized by X-ray diffraction (XRD) using a Rigaku diffractometer with Cu K$\alpha$ radiation. The detailed characterization of the samples is presented in Ref. [*4*]. The thickness of the VO$_2$ films grown on both types of substrate was 80nm.

The Raman spectra were obtained with a Kaiser Raman Rxn1 microprobe using a 50x magnification objective that provided spatial resolution of 1μm, and a depth of focus of 2μm. This instrument is configured to operate at three different wavelengths for the optical excitation: 532nm (25mW), 632.8nm (3.5mW), and 785nm (7.58mW) allowing us to compare the VO$_2$ Raman signal at these different optical wavelengths. It was also possible to use unpolarized incident beams, or to have the polarization of the incident laser beam linearly polarized in either the x or y direction (considering the normal of our sample to be the z-axis). In addition, we were able to set the polarization of the Raman signal to be either parallel or orthogonal to the incident pump polarization direction.

To control the temperature of the sample across the Insulator-Metal transition, we used a Peltier heater. A miniature 10k thermistor, pressed down with a thermal conducting paste on the surface of the sample, monitored the film temperature. We heated each sample, taking periodic measurements of the Raman spectra until the spectra stabilized as the phase transition was completed and the VO$_2$ peaks completely disappeared, at which point we would cool down the sample.

III. Effects of the laser wavelength and polarization in monoclinic VO$_2$ Raman studies

The experimental Raman spectra measured at low-temperature (monoclinic) for VO$_2$ samples grown on quartz and sapphire substrates are shown in Fig. 1. Group theory predicts a total of eighteen Raman active phonon modes in the low-temperature (insulating) regime: nine A$_g$ modes and nine B$_g$ phonon modes, that have been identified in previous works [13-15]. In our samples we have observed a majority of these identified modes (summarized in Table 1). It is important to note that we found none of the peaks associated with other forms of vanadium oxides, in particular, V$_2$O$_5$ that has similar Raman signature. That fact offers good evidence of the high purity of the VO$_2$ phase in our samples.

Fig. 1 shows the peaks of the VO2 film on quartz and sapphire. There are twelve clear $VO_2$ peaks identified by vertical lines: 198 $cm^{-1}$, 225 $cm^{-1}$, 265 $cm^{-1}$, 309 $cm^{-1}$, 342 $cm^{-1}$, 439 $cm^{-1}$, 481 $cm^{-1}$, 501 $cm^{-1}$, 595 $cm^{-1}$, 617 $cm^{-1}$, 660 $cm^{-1}$, and 810 $cm^{-1}$.

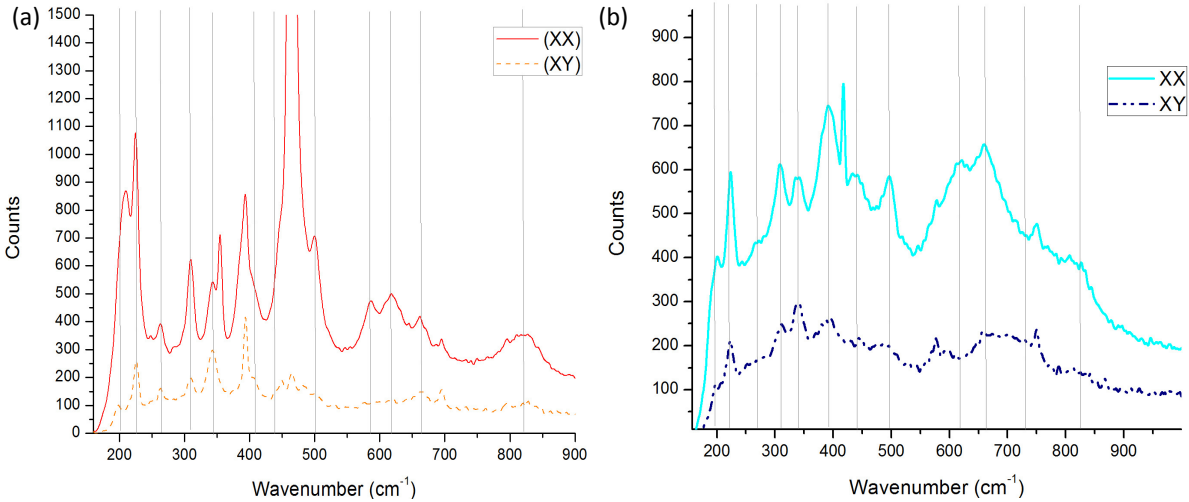

**Figure 1** Raman spectra for $VO_2$ thin films deposited on (a) quartz and (b) sapphire substrates. The Raman signal is detected in the same (XX) and in the orthogonal (XY)

Observation of some of the expected peaks is obscured because the $VO_2$ film is so thin (80nm) and overwhelmed by peaks from the substrate. In particular, for the film on $SiO_2$, the peak at 210 $cm^{-1}$ obscures the 194 $cm^{-1}$ $VO_2$ peak, and there is a large peak at 468 $cm^{-1}$ that obscures several $VO_2$ peaks. Those peaks are easier to see in the cross polarization configuration, since the $SiO_2$ peaks are attenuated. The $VO_2$ thin film grown on sapphire substrate shows additional peaks at 395 $cm^{-1}$ and 453 $cm^{-1}$, while 595 cm^-2 is blocked.

We also briefly investigated the dependence of the Raman signal on the polarization of the incident laser light. We found no noticeable difference between the Raman spectra for two orthogonal polarizations of the incident laser beam. We also compared the cases when the detected polarization of Raman signal was parallel or orthogonal to the excitation laser. We found no difference between Raman spectral peaks recorded in these two cases, even though in the cross-polarized measurements were noticeably weaker. These measurements are consistent with a transverse anisotropy of the $VO_2$ monoclinic lattice. However, polarization control may be a useful practical tool to minimize the presence of particular substrate peaks that can overwhelm the $VO_2$ peaks for substrates with polarization-dependent Raman response. The sample on the quartz substrate [Fig. 1(a)] can serve as an illustration: since quartz has distinctly different Raman signals for two polarizations of the detector, some $VO_2$ modes were better resolved in parallel polarization, and some in cross-polarization configuration, depending on what quartz peaks are suppressed (such as the large $SiO_2$ peak at 468 cm$^{-1}$). On the other hand, since sapphire has a triagonal symmetric lattice, its Raman spectrum is polarization-independent.

Figure 2(a) also shows strong dependence of the visibility of the $VO_2$ Raman resonances on the excitation light wavelength. Notably, the clearest spectrum with maximum number of resolved resonances was observed with near-IR excitation (wavelength 785 nm). Only about half of the peaks were detectable with a red He-Ne laser ($\lambda$=633 nm). For even shorter wavelength ($\lambda$=532 nm), the only visible Raman features were the substrate (quartz) Raman peaks. The $VO_2$ films grown on sapphire displayed even stronger wavelength sensitivity, illustrated in Fig 2(b). Only the infrared 785nm laser yielded an identifiable $VO_2$ spectrum, and we were not able to observe any Raman peaks due to the $VO_2$ thin film with either red or green lasers.

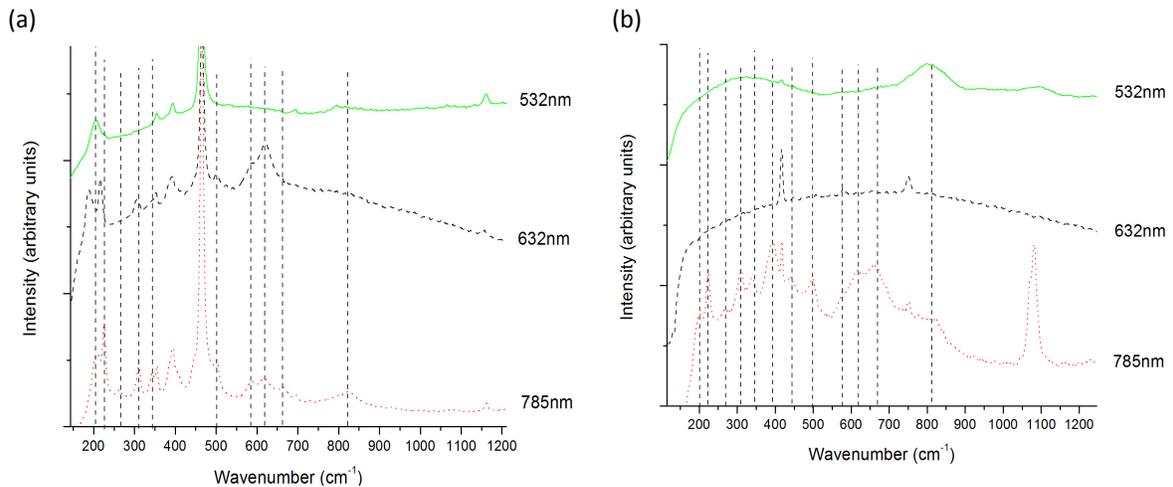

**Figure 2** Raman spectra for $VO_2$ thin films deposited on (a) quartz and (b) sapphire substrates obtained using different excitation lasers.

Such frequency dependence can be explained by increased contribution of the resonance fluorescence for higher-frequency radiation which increases the background and overwhelms any thin film contributions to the detected signal. Thus, our experiments suggest that it is advantageous to use a lower-frequency optical wavelength to probe the Raman structure of $VO_2$ thin films.

| Shilbe [13] | | Mei Pan [14] | | Chen [15] | | Present work | |
|---|---|---|---|---|---|---|---|
| Peaks | Modes | Peaks | Modes | Peaks | Modes | on Quartz | on Sapphire |
| 149 | | | | | | | |
| 199 | $A_g$ | 194 | $A_g$ | | | 198 | 200 |
| 225 | $A_g$ | 225 | $A_g$ | 226 | $A_g$ | 225 | 223 |
| 259 | $B_g$ | 258 | $B_g$ | 262 | $A_g$ | 250? | no |
| 265 | $B_g$ | 265 | $B_g$ | 264 | $B_g$ | 262 | 266 |
| 313 | $A_g$ | 308 | $A_g$ | 311 | $B_g$ and $A_g$ | 309 | 308 |
| 339 | $B_g$ | 339 | $B_g$ | 339 | $A_g$ | 342 | 339 |
| 392 | $A_g$ | 392 | $A_g$ | 390 | $A_g$ | blocked | unresolved |
| 395 | $B_g$ | 395 | $B_g$ | 395 | $B_g$ | blocked | 395 |
| 444 | $B_g$ | 444 | $B_g$ | 444 | $B_g$ | 439? | 443 |
| 453 | $B_g$ | 453 | $B_g$ | 454 | $B_g$ | blocked | 453 |
| 489 | $B_g$ | 489 | $B_g$ | 483 | $B_g$ | 481 | 486? |
| 503 | $A_g$ | 503 | $A_g$ | 500 | $A_g$ | 501 | 497 |
| 595 | $A_g$ | 585 | $A_g$ | 591 | $B_g$ | 595 | - |
| 618 | $A_g$ | 618 | $A_g$ | 618 | $A_g$ | 617 | 616 |
| 670 | $B_g$ | 650 | $B_g$ | 662 | $B_g$ | 662 | 660 |
| 830 | $B_g$ | 825 | $B_g$ | 826 | $B_g$ | 810 | 825 |

**Table 1 Comparison of measured monoclinic $VO_2$ Raman resonances with previously reported results.**

It is interesting to note that we have also observed several extra Raman peaks on the film on sapphire including one near 1075? $cm^{-1}$, that exhibits a temperature dependence consistent with the MIT transition. However, the frequencies of these peaks do not match any of previously identified photon modes in $VO_2$, and they lay beyond the expected Raman band for this material. Those peaks will require further study.

IV.     Raman studies of MIT in $VO_2$ thin films

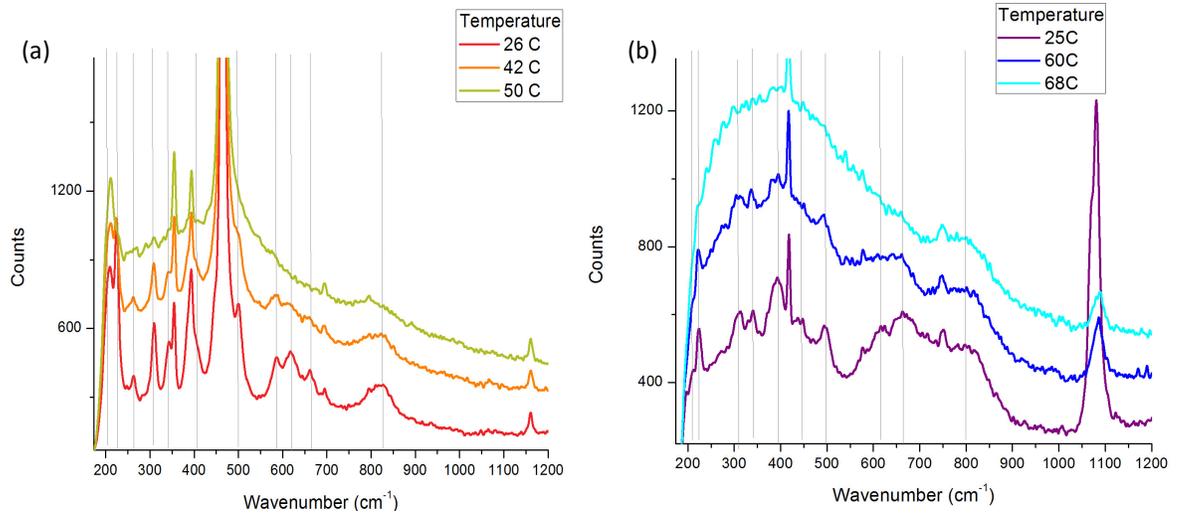

**Figure 3** Evolution of Raman spectra for VO$_2$ thin films deposited on (a) quartz and (b) sapphire substrates through thermally-induced IMT.

Figure 3 shows the sequence of Raman spectra taken as the VO$_2$ films were heated through the insulator-metal transition. As the lattice structure of the VO2 film changed, we observed the sharp spectral features associate with the monoclinic phase disappear, and get replaced by a spectrally broad features associated with the tetragonal conducting phase. On the other hand, the peaks from the substrates stayed unattenuated as the samples were heated, providing a suitable tool to identify Raman peaks due to each material. It is interesting to note that in previous work [14] the phonon modes of different symmetry groups display different temperature behavior. However, in our experiment we observed no such difference. In particular, a 224 cm$^{-1}$ peak (A$_g$ mode) and 660 cm$^{-1}$ (B$_g$ mode) disappeared at the same rate for both samples when the films were heated.

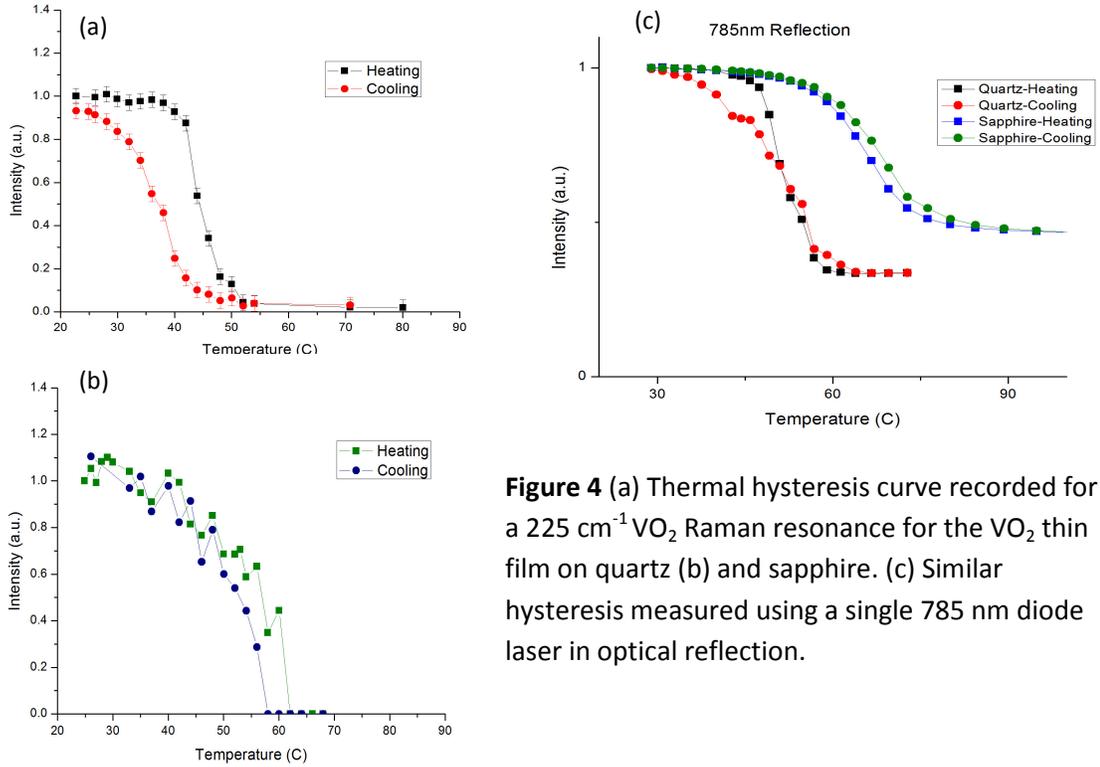

**Figure 4** (a) Thermal hysteresis curve recorded for a 225 cm$^{-1}$ VO$_2$ Raman resonance for the VO$_2$ thin film on quartz (b) and sapphire. (c) Similar hysteresis measured using a single 785 nm diode laser in optical reflection.

We note here that the VO$_2$ bulk transition temperature from insulator to a metal is 68 °C. The VO$_2$ peaks of the sample grown on sapphire disappear around the bulk transition temperature. However, we found that films grown on quartz exhibited a much lower transition temperature, around 50 °C. For more detailed transition analysis, we used a Gaussian fit function to measure the amplitude of the 224 cm$^{-1}$. Figure 4 (a) shows the dependences of the resonance amplitudes for this peak at on both quartz and sapphire. For the quartz substrate, the peaks disappear around 50 °C. For the sapphire substrate the peaks don't disappear until above the bulk transition temperature, 68 °C. The height of the peaks of the VO$_2$ on quartz shows a steep decline at the transition temperature; when plotted as a function of temperature, we obtain a typical hysteresis as we heat and cool the sample. The amplitude of the peaks on sapphire attenuate more slowly; while the change in reflectivity does happen at a higher temperature on the sample grown on sapphire, the overall change across the transition happens at much the same temperature range.

We also observed the increase in the background due to the broadband Raman signal from the VO$_2$ metallic phase, as illustrated in Figure 5. The VO$_2$ films on both substrates show a large increase in background as they transition to the metallic phase.

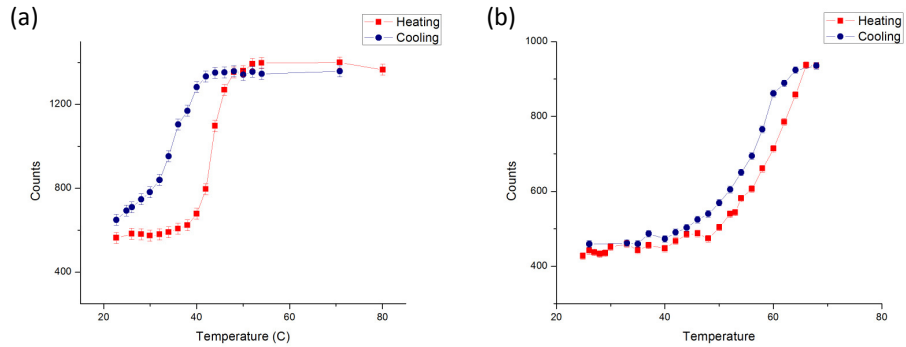

**Figure 5** Broadband Raman background measured around 225 cm$^{-1}$ for VO$_2$ thin films deposited on (a) quartz and (b) sapphire substrates.

It is also important to note that we have observed similar difference in the transition temperature by measuring the change in optical reflection of a 785 nm laser. It is known that for this optical wavelength the MIT leads to reduced reflection in the metallic phase, as shown in Fig. 4(b). However, for the VO$_2$ film on the quartz substrate the transition occurs at a much lower temperature (around 50$^o$C). The sapphire sample, on the other hand, displays the transition at higher temperature (around 68$^o$C), closer to the bulk transition temperature. In either case, these measurements are in agreement with Raman measurements.

Differences in behavior in the optical properties of VO$_2$ thin films deposited under identical conditions on different substrates in our case can be attributed to differences in the microstructure of the film. XRD measurements indicated that VO$_2$ films deposited on crystalline sapphire substrate tend to grow with a single crystal orientation, while films on crystalline quartz are polycrystalline, consisting on crystalline micrograins of VO$_2$ of different orientations. Previous experiments have demonstrated that doping [6] and the size of the nanostructures [14,17] can alter the average temperature of the phase transition, although the dominant effect in this case is broadening of the hysteresis loop. In our case, however, both samples display a very narrow width of thermal hysteresis, as seen in Fig. 4, indicating excellent homogeneity of each type of sample.

One possible explanation for the lower transition temperature in quartz $VO_2$ film is in the variation of the $VO_2$ film structure for different substrates. X-ray diffraction (XRD) analysis demonstrated that the film grown on quartz was textured along the out-of-plane direction, as shown in Fig. 6(a). The observed peaks were most likely M-$VO_2$(011). However, it had no preferred orientation for in-plane direction [Fig. 6(b)]. In contrast, $VO_2$ film grown on c-sapphire substrate is single phase [4]. Such difference in structure may affect the transition dynamics, in particular the formation of nanoscale transitional "puddles" inside the insulating phase as precursors MIT [18]. It is possible that early nucleation of such metal puddles is favored at grain boundaries in polycrystalline $VO_2$ films on quartz substrate, compared to the monocrystalline films grown on sapphire substrate. Similar behavior has been observed during magnetization reversal measurements on epitaxial vs. polycrystalline magnetic thin films of the same material, where earlier nucleation of reversed domains is favored at grain boundaries in the polycrystalline case [20].

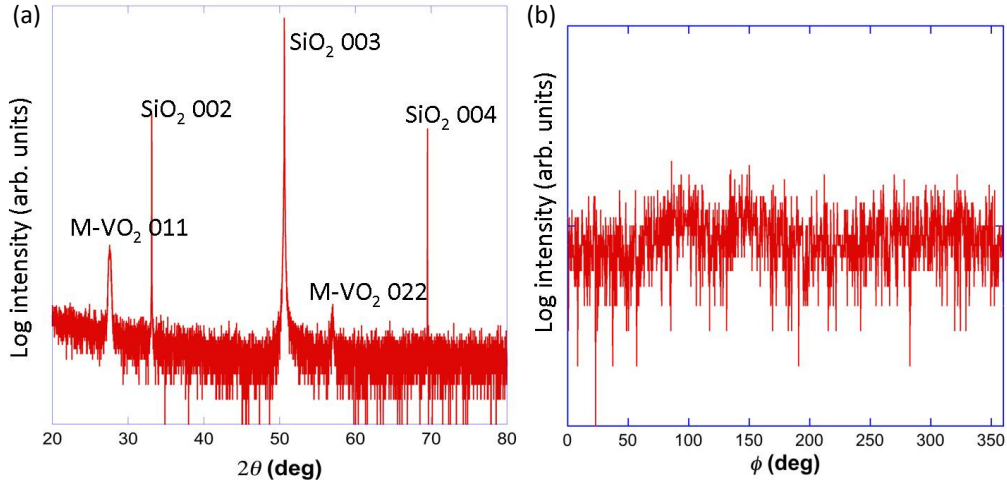

**Figure 6** (a) An XRD out-of-plane $\theta-2\theta$ scan for VO$_2$ thin films on quartz (SO$_2$) substrate, showing mostly a single phase M-VO$_2$ (011). (b) An XRD in-plane $\phi$ scan of the (011) plane; absence of detected peaks indicates a polycrystalline structure.

IV.     Conclusions

In conclusion, we were able to observe most of the previously identified VO$_2$ peaks and some new ones that warrant further study, on VO$_2$ films deposited on different substrates using Raman spectroscopy. The best spectral resolution was obtained with 785nm laser excitation. We observed that the peaks for the monoclinic VO$_2$ insulating phase deposited on different substrates disappear at different temperatures, indicating strong influence of the substrate on the MIT dynamics due to the substrate effect on the films microstructure. This difference in transition temperature was confirmed using optical reflection measurements. Our results are important for photonic applications of VO$_2$ thin films, since they show a venue to tailor the transition temperature via control of the crystalline microstructure employing different substrates. Further, our studies also showcase a suitable approach to investigate Raman signature peaks on thin films via controlling the response of the substrate; in the present case we show how polarization control may be used to minimize the presence of particular substrate peaks that can overwhelm thin film peaks for substrates with polarization-dependent Raman response.

This work is financed by NSF, DMR-1006013: Plasmon Resonances and Metal Insulator Transitions in Highly Correlated Thin Film Systems. We also acknowledge support from the NRI/SRC sponsored ViNC center and the Commonwealth of Virginia through the Virginia Micro-Electronics Consortium (VMEC).